\documentclass[preprint,aps]{revtex4}
\usepackage{graphicx}
\usepackage{dcolumn}
\usepackage{ifthen}
\newcommand{\type}{ps}

\begin{document}

\title {Current-Voltage Curves for Molecular Junctions Computed Using
All-Electron Basis Sets }
\author{Charles W. Bauschlicher, Jr.}
\email{Charles.W.Bauschlicher@nasa.gov}
\author{John W. Lawson}
\email{John.W.Lawson@nasa.gov}
\affiliation{Mail Stop 230-3\\
Center for Nanotechnology \\
NASA Ames Research Center\\
Moffett Field, CA 94035 }

\begin{abstract}
{\normalsize
  We present current-voltage (I-V) curves computed using all-electron basis
sets on the conducting molecule.  The all-electron results are very similar
to previous results obtained using effective core potentials (ECP).
A hybrid integration scheme is used that keeps the all-electron calculations
cost competitive with respect to the ECP calculations.
By neglecting the coupling of states to the contacts below a fixed energy cutoff,
the density matrix for the core electrons can be evaluated analytically.
The full density matrix is formed by adding this core contribution to
the valence part that is evaluated numerically.
Expanding the definition of the core in the all-electron calculations
significantly reduces the computational effort and, up to biases of about 2~V, the results
are very similar to those obtained using more rigorous approaches.  The convergence of the 
I-V curves and transmission coefficients with respect to basis set is discussed.  The addition
of diffuse functions is critical in approaching basis set completeness.
}
\end{abstract}

\maketitle
\section{Introduction}
    Several programs\cite{xr1,t2,t3,t4,t5,t6,t7,t8,t9,t10} have been developed to compute current-voltage
(I-V) curves for molecular junctions.  One feature of these codes has been the use of
effective core potentials (ECPs) to describe both the molecular junction as well as the metal contacts.
Part of the motivation for using an ECP in the Green's function based code is due to the numerical integration
associated with computing the density matrix.  If one includes the core orbitals in the calculations,
the integration range is much larger.  Consider the well-studied case of benzene-1,4-dithiol between
two Au surfaces; the valence orbital energies range from -0.8 to -0.1~Hartrees, while including the
benzene-1,4-dithiol core orbitals increases the orbital energy range from -88. to -0.1~Hartrees. Clearly
the number of points required to achieve the same accuracy in an all-electron calculation is much larger than
in an ECP calculation.\par
     While one expects the core electrons to be rather inert, the use of ECPs always introduces some additional
uncertainties into the problem.  Ideally one would like to use an all-electron treatment but avoid
the very large increase in the computer time associated with the numerical integration.  In this paper,
we compare our previous ECP results with those obtained using an all-electron approach.
The calculations are made tractable by splitting
the energy integral for the density matrix into two regions, one for the core and one for
the valence states.  Assuming that the core does not couple to the contacts, which is implicit in
ECP calculations, results in an integral that can be done analytically.  The remaining valence
portion is done numerically as in ECP calculations.

     In an all-electron calculation, the demarcation between core states and valence states is not clearly defined.
In our first series of tests, the lower limit of the numerical valence integration is the same as in the ECP
calculations.  That is, the definition of the core electrons in the ECP and all-electron
calculations is the same.  However, since the density of states of Au is essentially zero below the
d-electron band ($E \approx -15eV$), this suggests that the lower limit of the integration can be
increased further, and in a second set of tests, the size of the core is expanded to this level.
In addition, above this cutoff,
the Au valence density of states is quite constant. This suggests that the use of a energy independent
constant self-energy might be valid for Au in this region.  As shown by Damle, Ghosh, and Datta\cite{t4}, the use of
a constant self-energy allows one to replace the numerical Gauss-Legrendre integration,
even in the valence region, with an analytic expression,
which dramatically reduces the computational effort.  In this paper, we
also compare this fully analytic approach with the Gauss-Legrendre numerical integration.
We use benzene-1,4-dithiol between two Au (111) surfaces for our tests.  \par

\section{Methods}
    We use the same basic approach as described in previous work, namely,
the I-V curves are computed using the self-consistent, non-equilibrium, Green's function approach as
implemented by Xue, Datta, and Ratner\cite{xr1,xr2,xue031,xue032,xueth}.
Unlike our earlier work, the calculation of the Green's function is now divided into two regions
\begin{equation}
\rho_{ij}=\int_{-\infty}^{E_c} G^<_{ij}(E) \frac{dE}{2\pi i} + \int_{E_c}^{\mu} G^<_{ij}(E) \frac{dE}{2\pi i},
\end{equation}
one for the core states and the other for the valence states.  In the integrals,
$G^<(E)$ is the ``lesser" Green's function, $E_c$ is an energy cutoff
that divides the core and valence regions, and $\mu$ is the chemical potential .  For non-current carrying states,
$G^<(E)=1/(ES-F-\Sigma(E))$ where $F$ is the Fock matrix for the junction and $\Sigma(E)$ is
the self-energy that represents the coupling to the contacts.  If $\Sigma(E)$ is zero in the
core region, as it is implicitly for an ECP calculation, then the first integral can be evaluated
analytically as an elementary complex contour integral in terms of the core eigenvectors of the
Fock matrix (computed using density functional theory).
In ``small-core" all-electron calculations, we set $E_c$ equal to the ECP cutoff, $E\approx -60.0~eV$,
and therefore, the core orbitals include the C 1s and the S 1s, 2s, and 2p orbitals.
These are the same core orbitals as in the ECP.  Note that in an analogous ECP calculation, the
first integral having this value of $E_c$ would always be zero.
In ``large-core" calculations, however, $E_c$
is set equal to -15.0~eV.  In the valence region, current carrying states have
$G^<(E)=G_R(E) \Gamma (E) G_R^{\dagger}(E)$ where $G_R$ is the regarded Green's function and
$\Gamma = i(\Sigma-\Sigma^{\dagger})$ is the coupling function.
In the second integral, therefore, the valence contribution
to the density matrix is computed numerically, following the integration procedure described in the work
of Xue, Datta, and Ratner\cite{xr1,xr2,xue031,xue032,xueth}.  The full density matrix is then formed
by combining the two parts.  In this way, the core orbitals are allowed to relax in response to the
changes in the valence density due to coupling with the metal and the applied electric field. \par
Additionally, as pointed out in by Damle et al. \cite{t4}, since the density of states of Au
is flat near the Fermi energy, $\Sigma(E)$ can be well-approximated by a constant.  This
permits the valence integral to be done analytically as well.  We compare this fully analytic approach
to our hybrid integration method for both all-electron and ECP calculations.\par
   In this work, we report results for both zero bias transmission functions as well as full I-V
characteristics.  The transmission function is calculated using the Landauer equation
$T(E)=Tr[\Gamma_R G \Gamma_L G^{\dagger}]$ where $\Gamma_R,\Gamma_L$ are for the right and left
contacts.  The current can then be evaluated as an integral of $T(E)$ in an energy window around
the Fermi level.  $$ I= \frac{2e}{h} \int_{-\infty}^{\infty} T(E)\times [f(E-\mu_l)-f(E-\mu_r)] dE, $$ where $f$
is the Fermi function.  The current is of direct interest since it corresponds to an experimentally
observable quantity.  The transmission spectrum while directly related to the current also contains
important microscopic information.  In this work, we consider a number of different basis sets for
both all-electron and ECP calculations.  As we will see, the I-V curves, but also especially the
transmission spectra can vary substantially, depending on the basis set.
In addition to providing additional insight into the results, calculation of the transmission coefficient
is much less computationally demanding than determining an entire I-V curve, and is therefore
ideal for comparing different basis sets.
\par
   The geometry is taken from previous work\cite{cwb1}, where the -SC$_6$H$_4$S- fragment is connected to
two Au(111) surfaces, with the $C_2$ axis of the molecular fragment is perpendicular to the surfaces
and the S atoms are placed above a three-fold hollow at a distance of 1.905~\AA\ above the Au surface.
The extended molecule is connected to the two bulk electrodes following the tight binding
approach described in previous work\cite{xr1,xr2,cwb1,cwb2}.  As in the previous work, the extended molecule
contained a -SC$_6$H$_4$S- molecular fragment along with six gold atoms from each metal surface.
The extended-molecule electronic structure calculations are based on density functional
theory (DFT), using the pure BPW91\cite{becke,pw91} functional.  The $\alpha$ and $\beta$ spin
densities are constrained to be equal in Au$_6$-S-C$_6$H$_4$-S-Au$_6$ extended molecule calculations.
The Au atoms are described using the Los Alamos 11 valence electrons effective
core potential\cite{lanl1} and a minimal basis set with the most diffuse s, p, and d primitives deleted.
Recent work\cite{cwbyx} using clusters with 21 Au for each contact and larger Au basis sets support the use
of the six Au atoms and the small minimal basis set.  \par
     In this work we use several all-electron  basis sets.  We describe four basis sets here,
and describe modifications to these sets below.  The first is the 6-31G set\cite{popleb} for the C, H and S atoms.  The second
is the 6-31G* basis set for C and H and the 6-31+G* set for S.  The third is the 6-311G* set for C and H and
the 6-311+G* set for S.  The fourth is
the cc-pVTZ basis set\cite{cc1} for C and H and the aug-cc-pVTZ\cite{cc3} for S.  For simplicity, we
denote the basis sets using the name of the set used to describe the S atom.
We compare the all-electron results with our previous work that used the compact effective core potential
(CEP) and the associated valence basis sets\cite{sbk84}.  The valence double zeta (VDZ) set has a
31G description of the atoms. The VDZ+P set adds d polarization functions to the C and S and a set of
diffuse functions to the S atom.  For C and S, the VTZ basis set is a 121G valence set.
The VTZ+P set adds diffuse functions to the S atom and polarization functions to  the C and S atoms.
The VTZ+(2df,2p) adds two d and one f polarization function to the C and S atoms and a set of diffuse sp
functions to the S\cite{popleb} atoms, while for hydrogen, the 6-311G set with two sets of
p polarization functions\cite{popleb} is used. \par
\par
   The I-V curves are computed using the Gauss-Legrendre integration are at 300~K unless otherwise noted.
The analytic approach is by definition performed at 0~K.
The electronic structure calculations are performed using the
Gaussian03 program system\cite{gaussian}.
All of the Green's function calculations are performed using the code described previously\cite{xr1,xr2,xue031,xue032,xueth}
that has been modified for the hybrid and analytic integration.
\par
\section{Results and Discussion}
\begin{figure}
\ifthenelse{\equal{\type}{ps}} 
{\includegraphics{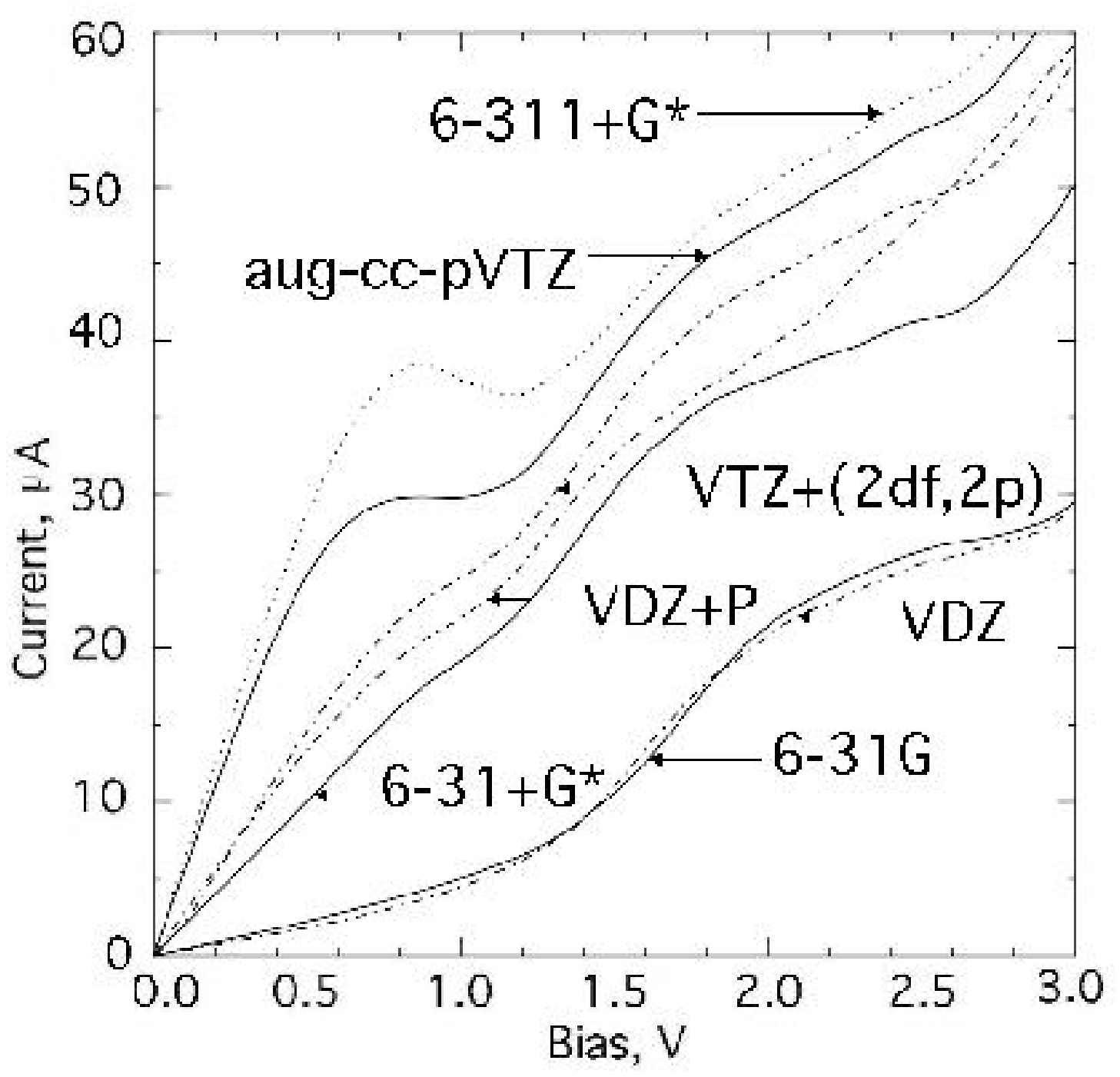}}
{\includegraphics{f1.pdf}}
\caption{\label{f1} The I-V curves for benzene-1,4-dithiol between two Au(111)
surfaces as a function of basis set used.  The dashed-dot curves are from ECP
calculations while the solid and dotted curves use an all-electron
treatment of the benzene-1,4-dithiol. }
\end{figure}

   The computed all-electron I-V curves are presented in Fig.~\ref{f1} along with three of our previous
I-V curves computed using ECPs.    The 6-31G set is double zeta in the valence region, and therefore
very similar in character to the VDZ basis set used in our earlier ECP calculations; an inspection  of
Fig.~\ref{f1} shows that I-V curves from these two treatments are very similar. Starting from
the 6-31G and VDZ sets and adding d polarization functions to the C and S atoms and  diffuse functions
to the S atom yields the 6-31+G* and VDZ+P sets, respectively.  An inspection
of Fig.~\ref{f1} shows that results obtained using the 6-31+G* and DZ+P basis sets are rather similar, but
adding the diffuse and polarization functions has increased the difference between the all-electron and
ECP basis sets, with biggest difference arising at bias values above 2.0~V.  \par
    In the ECP treatments~\cite{cwb1,cwb2}, improving the basis set from VDZ+P to VTZ+(2df,2p)
has only a small effect on the current.  The VTZ+P results, which are not shown, are similar to the VDZ+P
and VTZ+(2df,2p) results as expected.  Given the similarity of the VDZ+P and VTZ+P results,
it is a bit surprising to see significant differences between the 6-31+G* and 6-311+G* basis sets;
the initial slope with the 6-311+G* basis set is much larger than for the 6-31+G* set, and there
is a hump in the I-V curve at about 0.8~V.  Using the very large aug-cc-pVTZ basis set reduces the hump found
using the 6-311+G* basis set, but there is still a shoulder on the curve.  \par
\begin{figure}
\ifthenelse{\equal{\type}{ps}} 
{\includegraphics{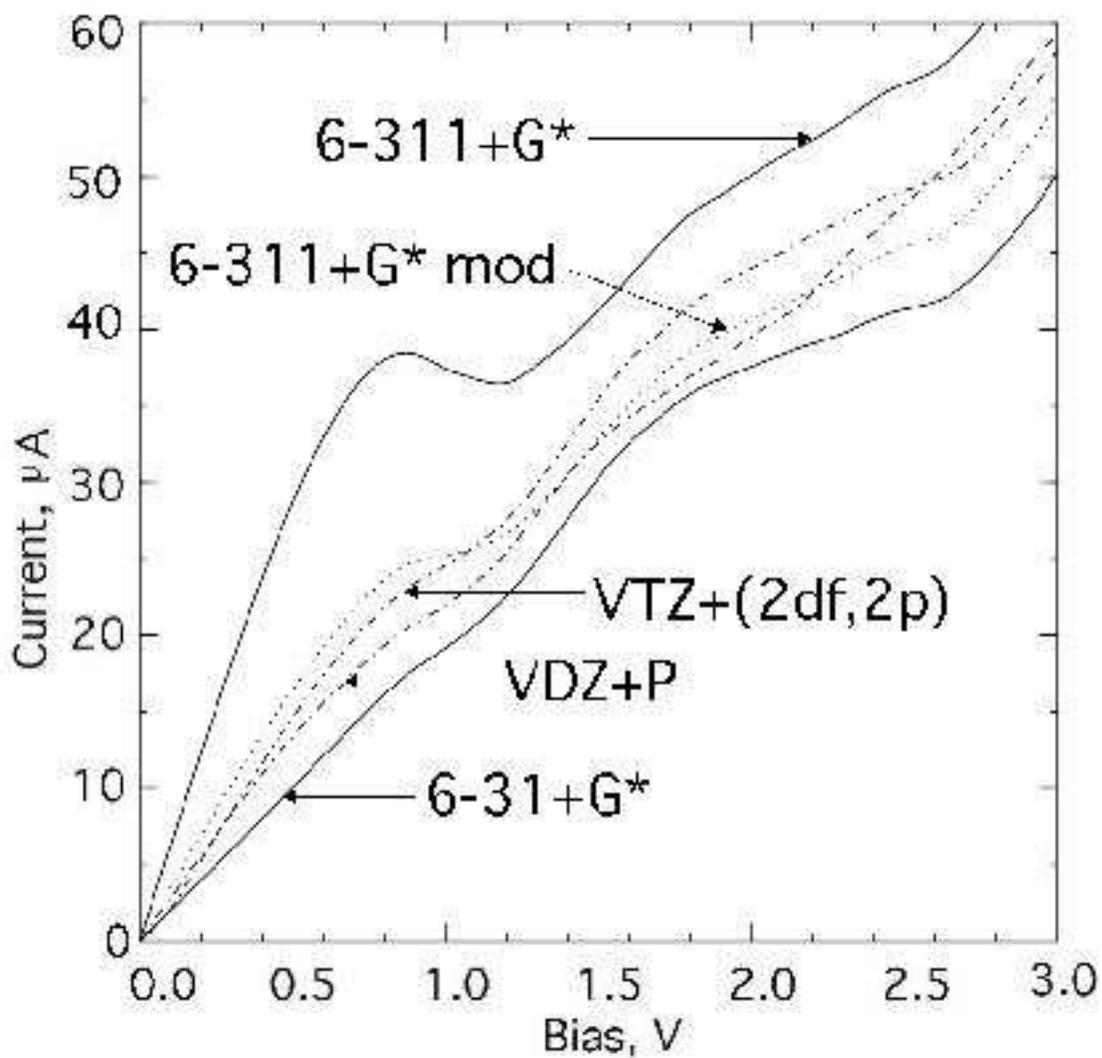}}
{\includegraphics{f2.pdf}}
\caption{\label{f2} Comparison of the 6-311+G* basis set with
the 6-311+G* basis set with the modified p function.  For
comparison the 6-31+G*, VDZ+P, and VTZ+(2df,2p) results
are also given. }
\end{figure}

   The I-V curves for the 6-311+G* and aug-cc-pVTZ basis sets appear nonphysical and therefore
they are investigated in more detail.  The most diffuse 2p functions of the 6-311+G* S
basis sets have exponents of 0.27746, 0.077141, and 0.0405.  The ratio between the first and second
is 3.56, while the ratio between
the second and third is 1.90.  The latter ratio is perhaps a bit small for a basis set that is not
approaching saturation and if we change the central exponent to 0.106, we make both ratios 2.62. Using
this modified basis set nearly eliminates the hump and yields an I-V curve that is  similar
to the 6-31+G* and ECP results for the VDZ+P and VTZ+(2df,2p) basis set, see Fig.~\ref{f2}.\par
\begin{figure}
\ifthenelse{\equal{\type}{ps}} 
{\includegraphics{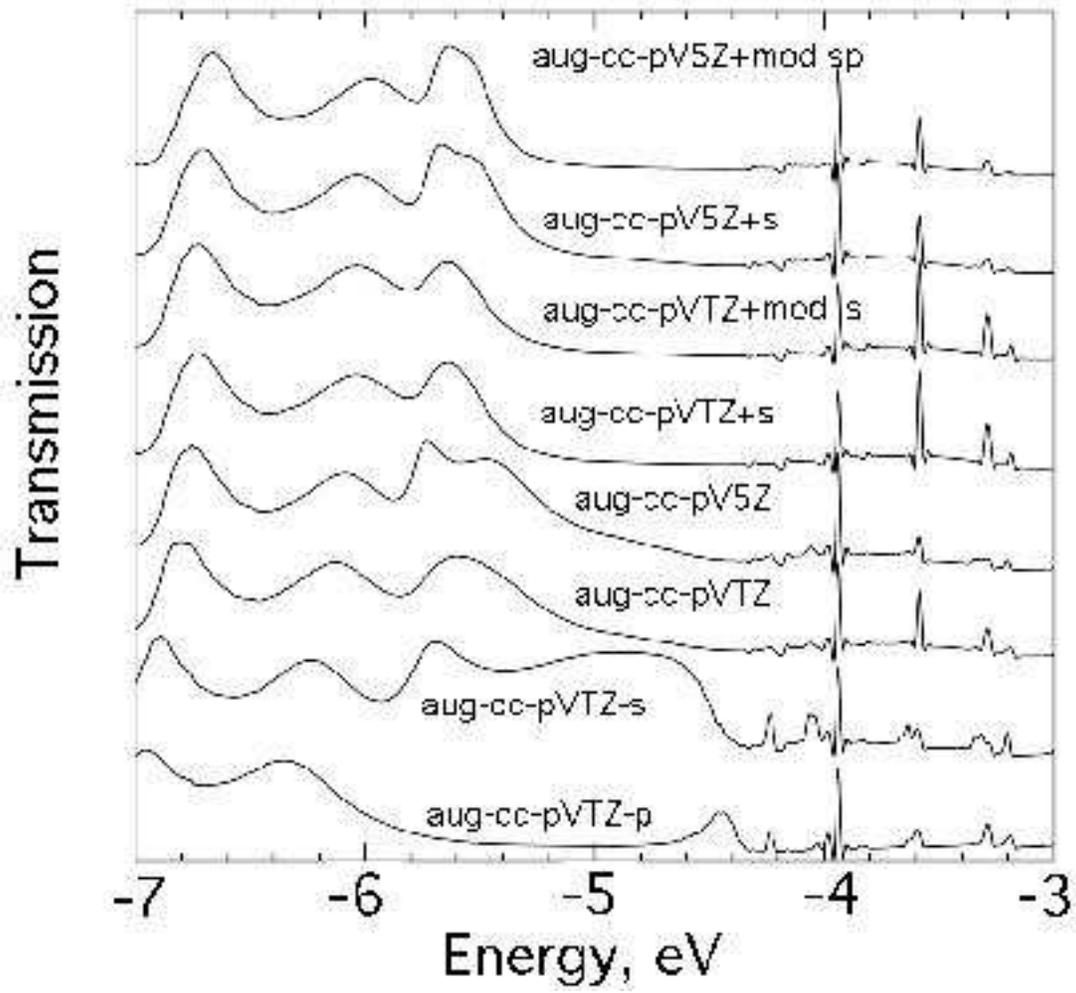}}
{\includegraphics{f3.pdf}}
\caption{\label{f3} Zero bias transmission coefficient for
several large all-electron basis sets. }
\end{figure}

   An inspection of the exponents of the aug-cc-pVTZ basis set does not show any obvious problems. To
get insight into the origin of the hump, we computed the transmission coefficient for the
aug-cc-pVTZ basis set and for the cases where we delete the most diffuse s or most diffuse p function,
see Fig.~\ref{f3}.  There are clearly dramatic changes in the transmission coefficient (especially
near -5.31 eV, which is the Fermi energy for Au)  when these functions are deleted.
The dramatic changes suggest that even the aug-cc-pVTZ basis set might not contain sufficient
diffuse functions.  To see if the S valence basis set is saturated, the sp set on S is replaced
with the aug-cc-pV5Z set and this makes only a very small change compared to the
aug-cc-pVTZ basis set.  This suggests that the
aug-cc-pVTZ set contains sufficient valence functions, and that problem probably lies in missing
diffuse functions. Adding one diffuse s (0.02) to the aug-cc-pVTZ basis set (labeled aug-cc-pVTZ+s
in the figure) makes a significant change in the transmission coefficient.
The aug-cc-pVTZ+mod s basis set starts from the aug-cc-pVTZ basis set and replaces the two most diffuse s
functions with 4 even tempered functions ($\beta$=2.2) (yielding a most diffuse s function of 0.014) and
yields a transmission curve that is very similar to the aug-cc-pVTZ+s set.  Adding a diffuse
s function (0.019) to the aug-cc-pV5Z set yield a tranmission curve that is very similar to
the aug-cc-pVTZ+s basis set. Our largest basis set starts from the aug-cc-pV5Z set an
removes the 5 most diffuse s and 5 most diffuse p functions and replaces them with 8 even-tempered
s functions and 7 even-tempered p functions ($\beta$=2.0); this results in the most diffuse s and p functions
having exponents of 0.01 and 0.0194, respectively.  This aug-cc-pV5Z-mod sp set is saturated in both the valence s and
p spaces and its transmission curve is very similar to the aug-cc-pVTZ+s, aug-cc-pVTZ-mod s, and
aug-cc-pV5Z+s sets.  This suggests that these large basis sets are essentially converged for the transmission
coefficient at zero bias. \par
\begin{figure}
\ifthenelse{\equal{\type}{ps}} 
{\includegraphics{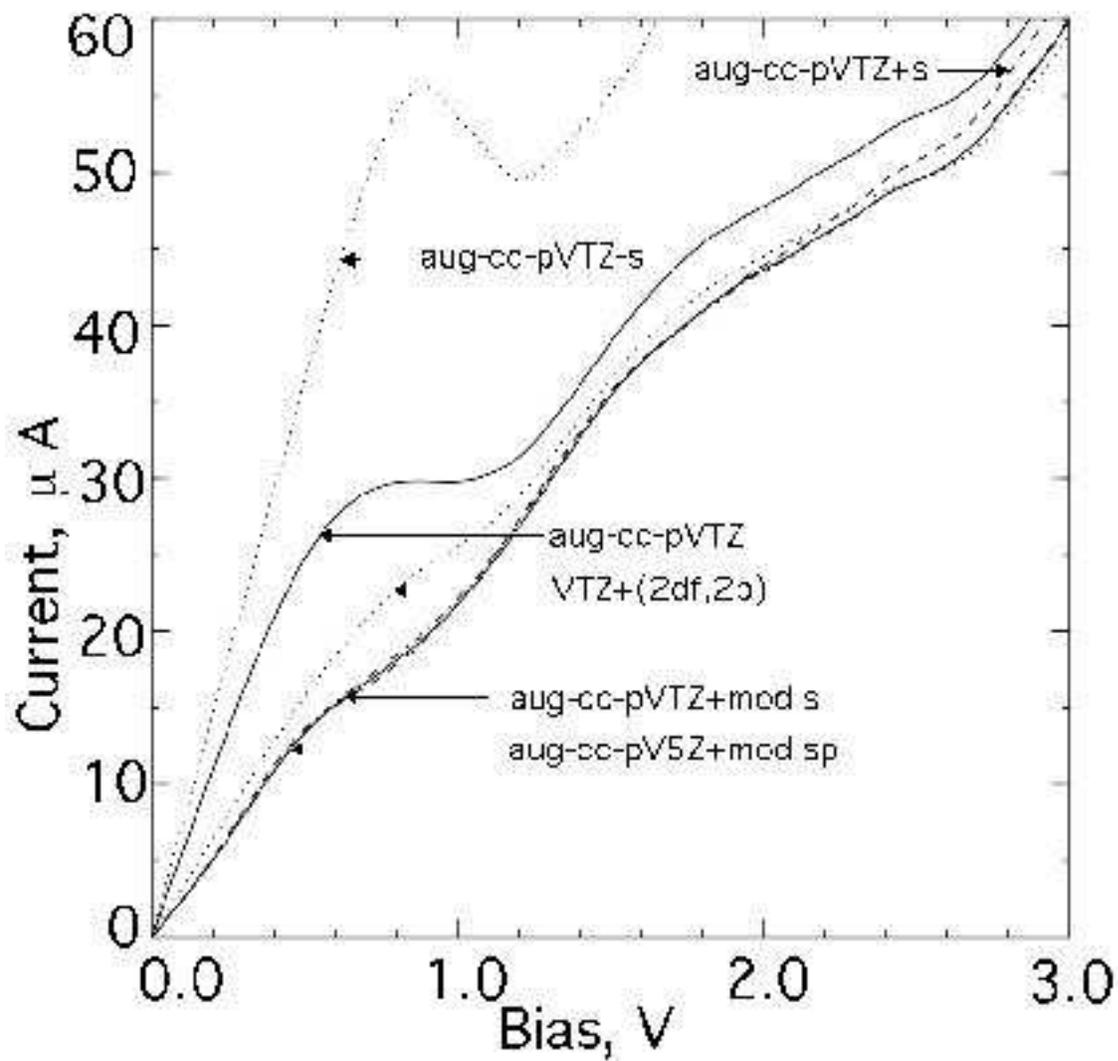}}
{\includegraphics{f4.pdf}}
\caption{\label{f4} The I-V curves for the all-electron treatments
with the largest basis sets.  The ECP results obtained using
the VTZ+(2df,2p) basis set are given
for comparison. }
\end{figure}

   For non-zero bias, we consider in Fig.~\ref{f4} the I-V curves for several of the large basis sets shown in Fig.~\ref{f3}
along with the ECP/VTZ+(2df,2p) set.  The aug-cc-pVTZ set has the hump as previously noted and
removing the diffuse s function set actually increases the current and the nonphysical hump, which is consistent
with the larger transmission near the Fermi energy.  The aug-cc-pVTZ+s, aug-cc-pVTZ+mod s, and aug-cc-pV5Z+mod sp
sets yield essentially the same curves till 2~V, where a small difference arises between the aug-cc-pVTZ+s and the
two sets with the saturated s or sp spaces.  The three large sets yield I-V curves that are very similar to the
VTZ+2df, which was the largest valence set considered
for the ECP approach.  \par
\begin{figure}
\ifthenelse{\equal{\type}{ps}} 
{\includegraphics{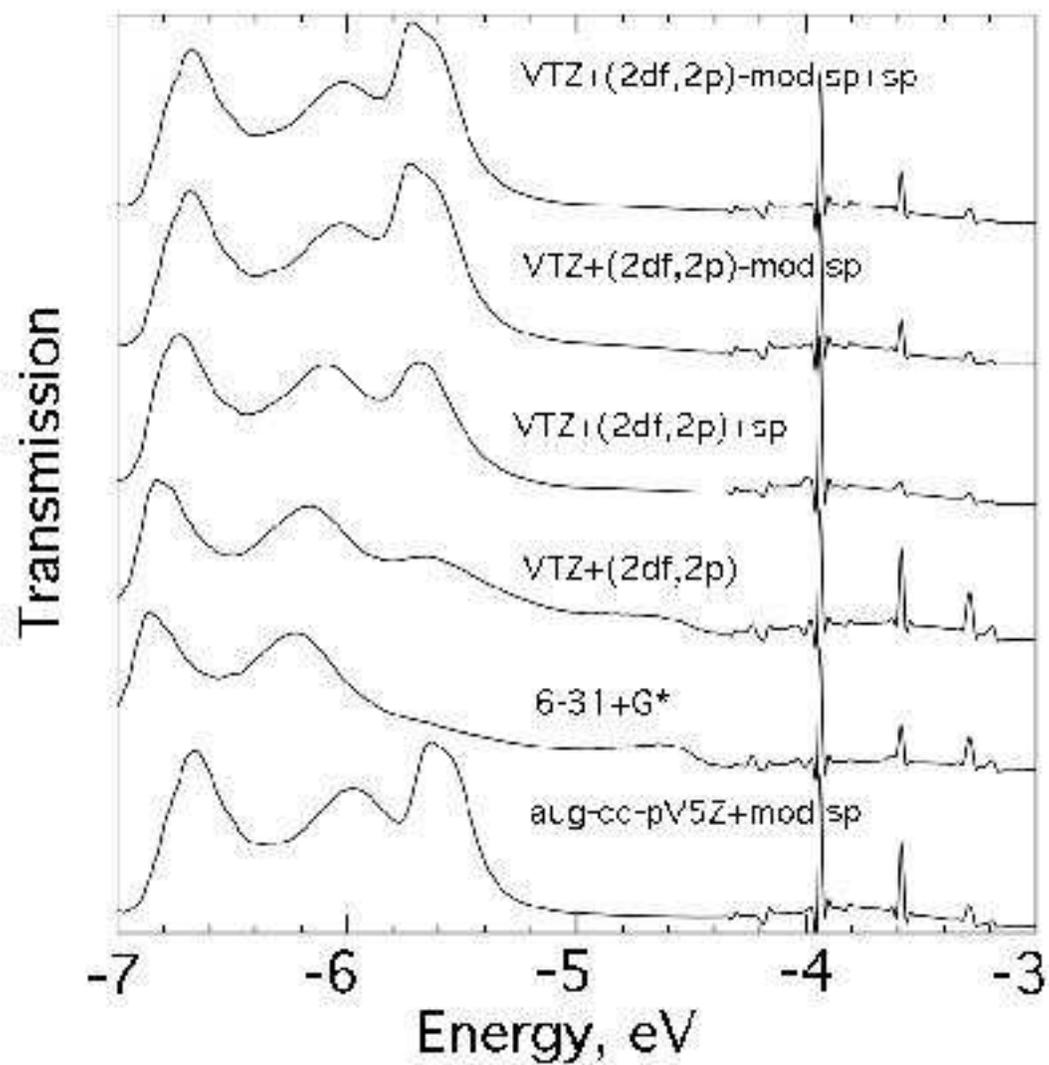}}
{\includegraphics{f5.pdf}}
\caption{\label{f5} Zero bias transmission coefficient for
several large ECP basis sets.  The largest all-electron
basis set results are given for comparison. }
\end{figure}

   We compare the transmission coefficients from the largest all-electron basis set with the 6-31+G* and
VTZ+(2df,2p) sets in Fig.~\ref{f5}.  The difference are much larger than might have been assumed from an
inspection of the I-V curves.  We therefore compute the transmission coefficients for some variants of the
VTZ+(2df,2p) basis set.  Adding a second set of diffuse sp functions (0.02) yields a transmission
coefficient that is very similar to the largest all-electron basis set.  Replacing the outermost
two sp functions in the VTZ+2df set with 4 even-tempered ($\beta$=2.0, the most diffuse
exponent is 0.018, and the set is denoted VTZ+2df-mod sp) yields a curve that is
essentially the same as the largest all-electron set.  Adding
one additional set of diffuse  (0.009) sp functions to the
VTZ+2df-mod sp set makes essentially no additional changes
in the transmission coefficient, showing that the VTZ+2df-mod sp is essentially converged with
respect to diffuse functions.\par
\begin{figure}
\ifthenelse{\equal{\type}{ps}} 
{\includegraphics{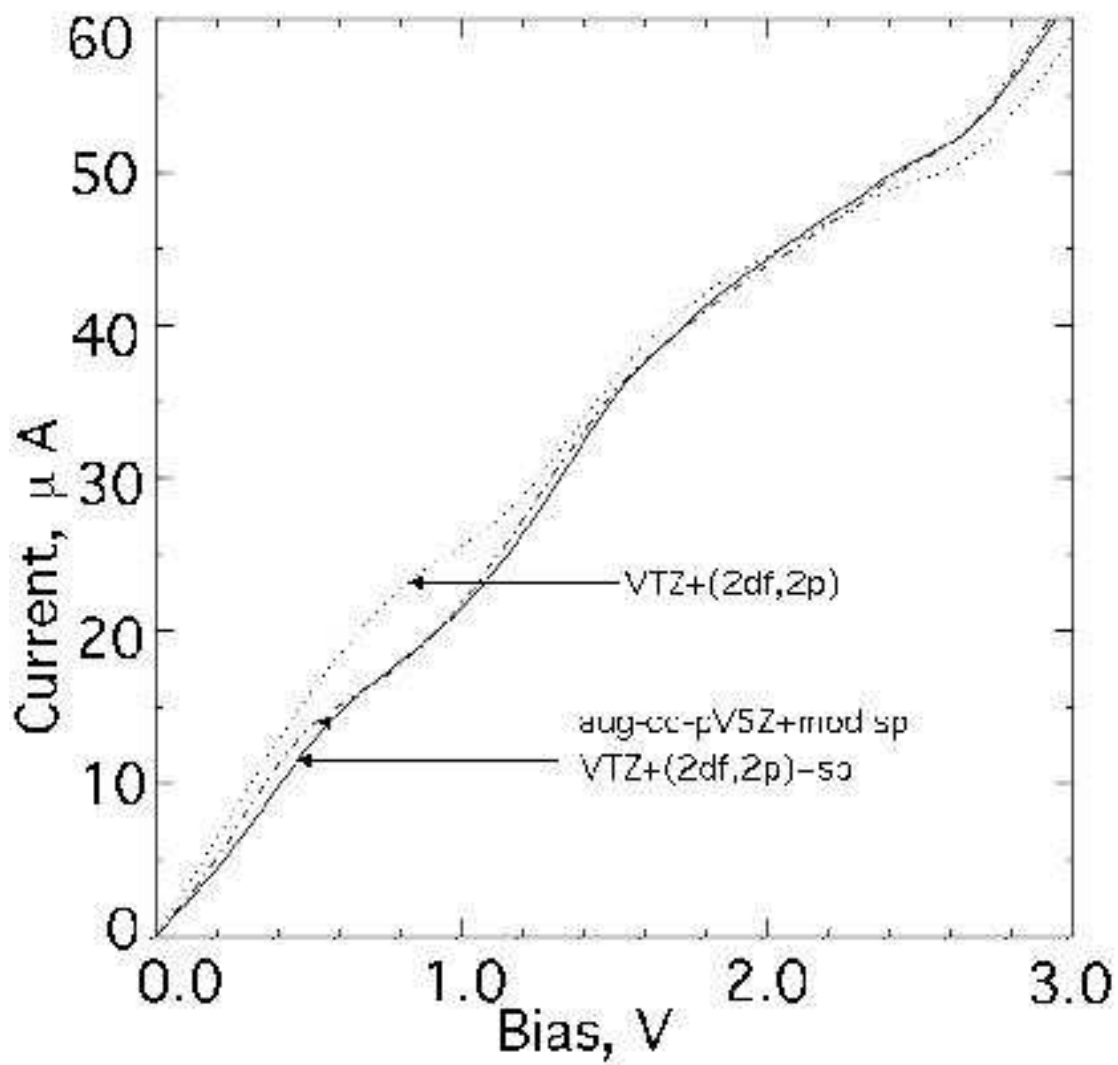}}
{\includegraphics{f6.pdf}}
\caption{\label{f6} The I-V curves for the large ECP basis
set treatments.  The largest all-electron
basis set results are given for comparison. }
\end{figure}

   In Fig.~\ref{f6}, we compare the I-V curve from  best all-electron calculation with the
VTZ+(2df,2p), VTZ+(2df,2p)+sp, and VTZ+(2df,2p)-mod sp+sp basis sets.   Note that
the VTZ+(2df,2p)-mod sp and VTZ+(2df,2p)-mod sp+sp basis sets yield very similar
I-V curves, and therefore the   VTZ+(2df,2p)-mod sp is not shown.
Adding the diffuse sp functions to the VTZ+(2df,2p) basis set brings the ECP and all-electron 
calculations into excellent agreement.  The VTZ+(2df,2p)-mod sp+sp basis set actually
agrees better with the best all-electron results till about 0.5 V, but then the VTZ+(2df,2p)+sp 
basis set actually agrees better until larger voltages, where all three curves are very similar.  All of
these large basis sets are in good agreement with the VTZ+(2df,2p).  Thus
our previous conclusion that the VTZ+(2df,2p) basis set was nearly converged for the I-V
curves for benzene-1,4-dithiol on Au(111) is valid, but the difference in the transmission
curves shows that for a metal with a different Fermi level, that conclusion might not have
been true.  On the basis of the transmission curve, additional diffuse function are required
to reach the basis set limit for this transmission coefficient, and therefore for the I-V curves
for and an arbitrary metal.  
 \par
   The calculations presented in this work show that the all-electron basis sets yield results that
are very similar to those obtained using the ECP approach in conjunction with similar quality
valence basis sets.  While the all-electron calculations lead to larger basis sets, the
number of iterations required to obtain convergence appears, on the average, to be smaller for
the all-electron treatment.  For example the 6-31+G* calculations require an average of about 10
iterations to converge for biases up to about 2~V, while the VTZ+P calculations require, on the
average, about 20 iterations.  Above 2~V, the number of iteration can vary significantly
with the bias voltage and it is a bit more difficult to make any conclusions.  For other
conducting molecules or for different treatments of the metal contacts, we have found
that the convergence of the all-electron calculations tends to be a bit better than
that found using the ECP.  Thus the hybrid all-electron calculation may offer computational
benefits that off set the larger number of basis functions.\par
\begin{figure}
\ifthenelse{\equal{\type}{ps}} 
{\includegraphics{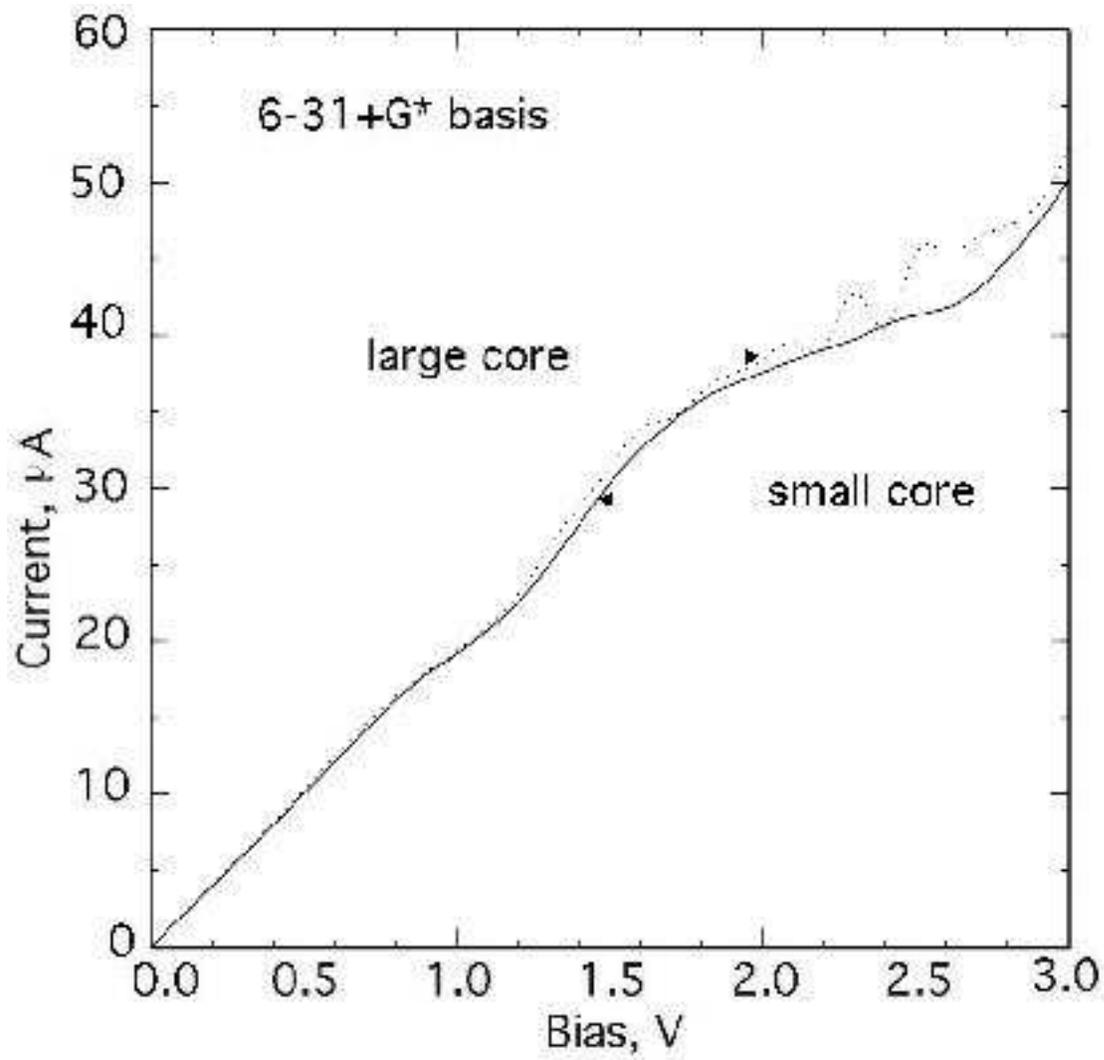}}
{\includegraphics{f7.pdf}}
\caption{\label{f7} The I-V curves for the all-electron treatments
with the small core and the large core.  Curves are almost identical
for $V<2.0$~V, but show small differences at higher bias.  We show
results for the basis set 6-31+G* which had the largest discrepancy
for $V>2.0$~V }
\end{figure}

    For the results presented above, the size of the core was picked to be the same as
in the ECP calculation.  However, we can include more electrons into the core if we
pick the integration cutoff based on the Au density of states, which results in
seven additional orbitals being treated as core; these include the S
3s orbitals and five of the six C 2s orbitals.  In addition to changing the cutoff,
since we are integrating over a shorter region, we reduce the number of integration
points by a factor of four.  A plot of the small core vs. large core results is shown
in Fig.~\ref{f7}, where up to 2~V, the two curves are nearly identical, but the
large core calculation require about one half the computer time as the small core calculations. Above 2~V
there are some differences, with the large core curve showing some wiggles.  The
difference between the large core and small core varies with basis set, with the
differences being the largest for the 6-31+G* basis set, which is shown in Fig.~\ref{f7}.
While more tests are required, it appears that the use of the larger core offers
a reduced computational cost, with only small changes in the quality of the
results, especially at lower voltages.  We should also note that this approach
can also be used in conjunction with the ECP basis set to reduce the integration
range and the number of integration points.  Finally we should note that
the choice of the cut off depends on the metal, and therefore this approach
should be tested before being used on metals other than Au.\par
\begin{figure}
\ifthenelse{\equal{\type}{ps}} 
{\includegraphics{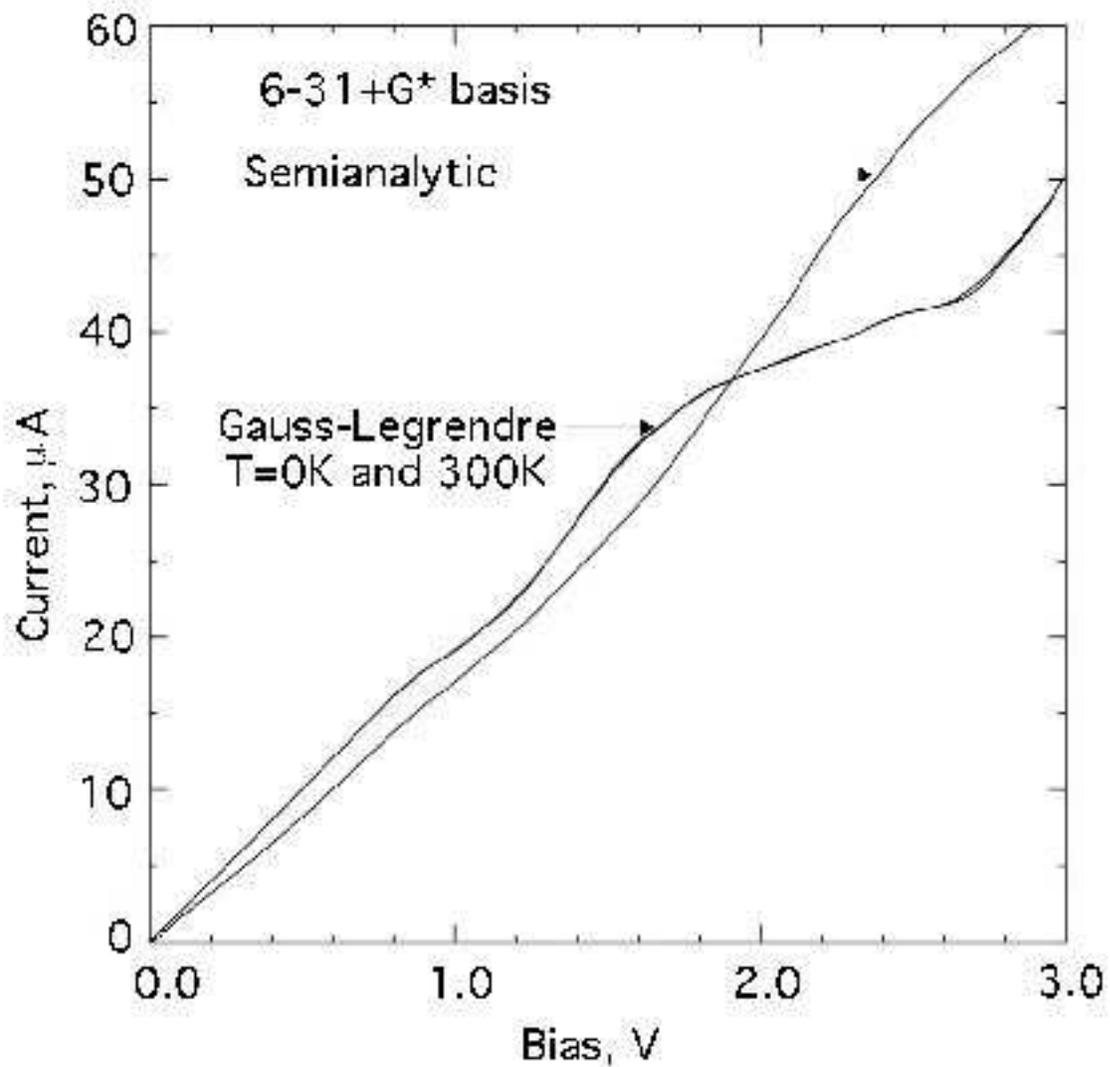}}
{\includegraphics{f8.pdf}}
\caption{\label{f8} A comparison of the results obtained using
the fully analytic integration of Damle, Ghosh, and Datta
and the Gauss-Legrendre approaches for the small core.
The analytic calculations are at 0~K. }
\end{figure}

    As noted by Datta and co-workers\cite{t4}, the density of states of Au near the Fermi level
is very flat and can be approximated by a constant.  If this is done, it is possible
to replace the Gauss-Legrendre integration with analytic expressions.  This leads
to a significant speed up in the evaluation of the new density matrix.  In Fig.~\ref{f8}
we show a comparison of the analytic approach with the Gauss-Legrendre integration
for the 6-31+G* all-electron basis set at 0 and 300~K.  The Gauss-Legrendre integration
results at 0 and 300~K are essentially identical, so any difference between the fully
analytic approach, which by definition is at 0~K, and the Gauss-Legrendre integration
results is not due to temperature.  Overall, the agreement between the analytic
and Gauss-Legrendre approaches is good up to biases of about 2~V. Thus for Au, the
analytic approach is probably a good approximation up to about 2~V.  We should
note that the analytic integration can also be used in conjunction with
the ECP basis set, and like the all-electron calculations, we find the analytic integration
to be in excellent agreement with the Gauss-Legrendre integration for biases below
about 2~V.
\par
\section{Conclusions}
   For benzene-1,4-dithiol between two Au(111) surfaces, the results obtained using
all-electron and ECP approaches yield similar results when  similar quality valence basis
sets are used.  The all-electron calculations using large basis sets suggest that
the Greens function approach used is more sensitive to basis set dependencies than
standard energy calculations.
For the limited number of test performed, the all-electron
calculations appear to converge faster than the ECP calculations.  For Au(111), expanding
the size of the core or changing to a analytic approach speed up the calculations
without degrading the results for biases below about 2~V.
\par
\section{Acknowledgments}
 C.W.B is a civil servant in the Space Technology Division (Mail Stop 230-3), while
J.W.L. is a civil servant in the IC Division (Mail Stop 269-2).\par

\end {document}